\documentclass[aps,prl,twocolumn]{revtex4}
\usepackage{amssymb}
\usepackage{graphicx}
\usepackage{amsmath}

\begin{document}

\title{Floquet Topological States in Shaking Optical Lattices}
\author{Wei Zheng and Hui Zhai}
\affiliation{Institute for Advanced Study, Tsinghua University, Beijing, 100084, China}
\date{\today }

\begin{abstract}
In this letter we propose realistic schemes to realize topologically
nontrivial Floquet states by shaking optical lattices, using both
one-dimension lattice and two-dimensional honeycomb lattice as examples. The
topological phase in the two-dimensional model exhibits quantum anomalous
Hall effect. The transition between topological trivial and nontrivial
states can be easily controlled by shaking frequency and amplitude. Our
schemes have two major advantages. First, both the static Hamiltonian and
the shaking scheme are sufficiently simple to implement. Secondly, it
requires relatively small shaking amplitude and therefore heating can be
minimized. These two advantages make our scheme much more practical.
\end{abstract}

\maketitle

Topological state of matters such as quantum Hall effect and topological
insulator have been extensively studied in equilibrium systems. Recently,
topological classification of quantum states in a periodically driven
non-equilibrium system has been proposed \cite{FTI,Demler}, in which the
topologically nontrivial states are named as \textquotedblleft Floquet
topological insulator" \cite{FTI}. Floquet topological band has been first
realized in photonic crystal and the edge state of light has been observed
\cite{Phonoic}. While so far it has not been realized in any solid-state or
cold-atom system.

Realizing and studying topological state of matter is also one of the major
treads for cold atom physics nowadays, for which Raman laser coupling \cite%
{Bloch,spielman,Bloch2013,Ketterle2013} and shaking optical lattice \cite%
{Sengstock1,Sengstock2, Cheng} have been developed as two major schemes. In
several recent experiments, it has been demonstrated that fast shaking
optical lattices can generate synthetic abelian gauge field and magnetic
flux \cite{Sengstock1,Sengstock2}. In this letter we propose that shaking
optical lattice is also a powerful tool to realize Floquet topological state
in cold atom systems.

We first demonstrate that in one-dimension lattice it realizes a system
equivalent to Su-Schrieffer-Heeger model \cite{SSH} with nonzero Zak phase;
Then, we show that in two-dimension honeycomb lattice \cite{Esslinger} it
realizes a system equivalent to the Haldane model which exhibits quantum
anomalous Hall effect \cite{Haldane}. So far, quantum anomalous Hall effect
has only been found in chromium-doped $\text{(Bi,Sb)}_2\text{Te}_3$, and
growing this material is extremely challenging \cite{AQH}. It is therefore
highly desirable that one can quantum simulate this effect with cold atom
system. However, despite of several proposals \cite{AQH_cold_atom} this
effect has not yet been successfully simulated in cold atom setup. Our
scheme has two major advantages for which it becomes much practical.

The first is its simplicity. To realize a topological state in a static
system, it usually requires particular form of hopping term. For instance,
in order to realize the Haldane model \cite{Haldane}, one needs to generate
a special next-nearest range hopping term, which usually requires
engineering laser-assisted tunneling in cold atom system \cite%
{Bloch,spielman,Bloch2013,Ketterle2013}. In contrast, in our scheme, the
static Hamiltonian is quite simple (it only contains normal nearest
neighboring hopping without extra phase factor) and has been realized in
different laboratories already. The beauty of this scheme is that such a
simple static Hamiltonian can result in a topological nontrivial state when
a proper shaking term is turn on, and such a shaking can also be very easily
implemented by time-periodically modulating the relative phase of two
counter-propagating laser beams.

The second is minimizing heating. In our proposals, it only demands quite
small shaking amplitude in order to reach topological phase, and
consequently heating is minimized. In fact, our proposals are inspired by
recent experiment in Chicago group, in which $s$-band and $p$-band of a
one-dimensional optical lattice is resonantly coupled by shaking. It is
shown that the band dispersion can be qualitatively changed even with small
shaking amplitude and the accompanying heating is insignificant \cite{Cheng}.

\textit{General Method. } Our theoretical treatment of shaking optical
lattices is based on the Floquet theory. For a periodical driven Hamiltonian
$\hat{H}(t)$ with period $T$, its Floquet operator can be defined as
\begin{equation}
\hat{F}=\hat{U}\left( T_{i}+T,T_{i}\right) =\hat{\mathcal{T}}\exp \left\{
-i\int_{T_{i}}^{T_{i}+T}dt\hat{H}\left( t\right) \right\} ,
\label{Floquet_operator}
\end{equation}%
where $\hat{\mathcal{T}}$ denotes time-order, and $T_{i}$ is the initial
time. The eigenvalue and eigenstates of $\hat{F}$ is given by
\begin{equation}
\hat{F}\left\vert \varphi _{n}\right\rangle =e^{-i\varepsilon
_{n}T}\left\vert \varphi _{n}\right\rangle,  \label{eigen}
\end{equation}%
where $-\pi /T<\varepsilon _{n}<\pi /T$ is the quasi-energy. In this work we
shall use two different methods listed below to show how simple shaking
schemes can result in nontrivial topology in optical lattice systems,
because each method has its own advantage.

\textbf{Method I.} We can numerically evaluate Floquet operator $\hat{F}$
according to Eq. \ref{Floquet_operator} and determine its eigenvalues and
eigen-wave-functions from Eq. \ref{eigen}. If a periodically driven system
exhibits nontrivial topological, there must be in-gap quasi-energies $%
\epsilon $ and their corresponding wave functions $\varphi $ are spatially
well localized at the edge of the system \cite{Demler}. The advantage of
this method is that once $\hat{H}(t)$ is given, it is free from any further
approximations.

\begin{figure}[tbp]
\includegraphics[width=2.8in]
{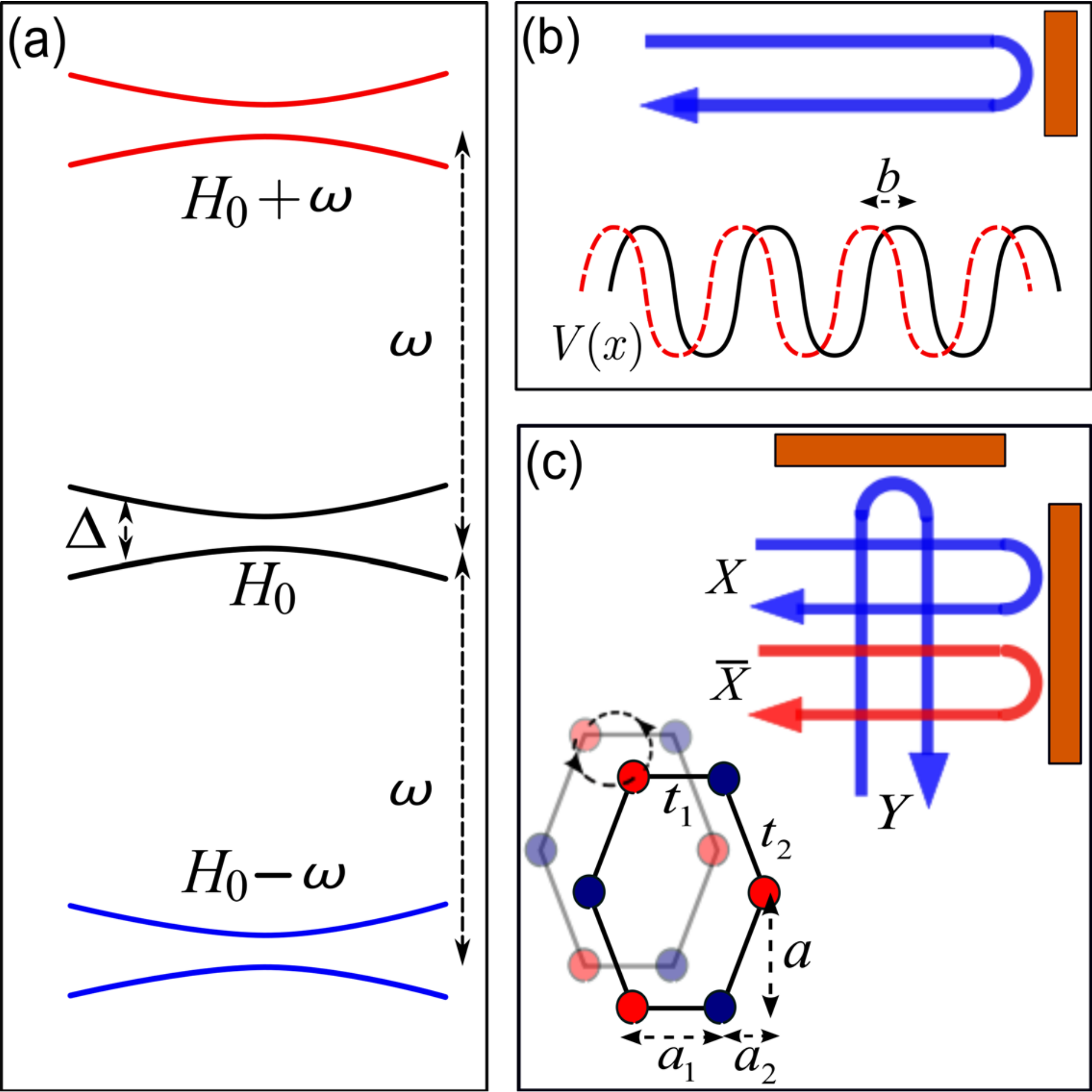}
\caption{ (a) The typical energy structure under consideration. (b) The
laser setup of the one-dimensional shaking optical lattice. Solid and dashed line represent lattice potential at two different time. (c) The laser
setup of the two-dimensional honeycomb optical lattice. The dashed circle with arrow indicates how lattice potential rotates in time. }
\label{fig1}
\end{figure}

\textbf{Method II.} We can introduce a time-independent effective
Hamiltonian $\hat{H}_{\text{eff}}$ via $\hat{F}=e^{-i\hat{H}_{\mathrm{eff}%
}T} $. Expanding $\hat{H}(t)$ as $\hat{H}(t)=\sum_{n=-\infty }^{\infty }\hat{%
H}_{n}(t)e^{in\omega t}$ with $\omega =2\pi /T$, we consider a situation as
shown in Fig.\ref{fig1}(a), that is, the static component $\hat{H}_{0}$
contains $m$-bands within an energy range of $\Delta $ and $\omega \gg
\Delta $. The two concrete examples discussed later either belong to this
situation or can be transferred into this situation by a rotating wave
transformation, with $m=2$. Under this condition, it is straightforward to
show that to the leading order of $\Delta /\omega $, $\hat{H}_{\text{eff}}$
can be deduced as
\begin{equation}
\hat{H}_{\mathrm{eff}}=\hat{H}_{0}+\sum\limits_{n=1}^{\infty }\left\{ \frac{%
\left[ \hat{H}_{n},\hat{H}_{-n}\right] }{n\omega }-\frac{\left[ \hat{H}_{n},%
\hat{H}_{0}\right] }{e^{-2\pi i\alpha }n\omega }+\frac{\left[ \hat{H}_{-n},%
\hat{H}_{0}\right] }{e^{2\pi i\alpha }n\omega }\right\} ,  \label{Heff}
\end{equation}%
where $T_{i}$ is taken as $\alpha T$ with $0\leq \alpha <1$. Since $\hat{F}$
with different choices of initial time $T_{i}$ relate to each other by a
unitary transformation, quasi-energy is independent of the choice of $T_{i}$%
. Practically, we can choose an optimal $\alpha $ that simplifies $\hat{H}_{%
\text{eff}}$. Then we can apply schemes developed for a time-independent
Hamiltonian to classify topology of $\hat{H}_{\text{eff}}$. Although this
method involves further approximations, it has the advantage that it is
physically more transparent and can bring out the connection to topological
phenomena in equilibrium systems.

\textit{One-Dimensional Case. } A one-dimensional lattice is formed by two
counter-propagating lasers. As one time-periodically modulates the relative
phase $\theta $ between two lasers, it will result in a time-dependent
lattice potential, as shown in Fig.\ref{fig1} (b),
\begin{equation}
H\left( t\right) =\frac{\hat{k}_{x}^{2}}{2m}+V\cos ^{2}\left[ k_{r}x+\theta
\left( t\right) \right]
\end{equation}%
where $\theta \left( t\right) =k_{r}b\cos \left( \omega t\right) $, and $b$
is the maximum lattice displacement. By transferring to the comoving frame, $%
x\rightarrow x+b\cos \left( \omega t\right) $, the Hamiltonian acquires a
time-dependent vector potential term as
\begin{equation}
H\left( t\right) =\frac{\hat{k}_{x}^{2}}{2m}+V\cos ^{2}\left( k_{r}x\right)
-b\omega \sin \left( \omega t\right) \cdot \hat{k}_{x}.
\end{equation}%
The first two static terms give a static band structure with Bloch wave
function $\varphi _{\lambda }(k_{x})$. In this bases, by only keeping $s$-
and $p$-band (for the reason which will be clear later), we can write down a
tight-binding Hamiltonian as
\begin{equation}
\hat{H}\left( t\right) =\sum\limits_{i}\hat{\Psi}_{i}^{\dag }K\left(
t\right) \hat{\Psi}_{i}+\sum\limits_{i}\left( \hat{\Psi}_{i}^{\dag }J\left(
t\right) \hat{\Psi}_{i+1}+h.c.\right)  \label{H1d}
\end{equation}%
where $\hat{\Psi}_{i}^{\dag }=(\hat{a}_{\text{p},i}^{\dag },\hat{a}_{\text{s}%
,i}^{\dag })$ are creation operators for $s$- and $p$-orbitals. And
\begin{align}
&K\left( t\right)=\left(
\begin{array}{cc}
\epsilon _{\text{p}} & ih_{0}^{\text{sp}}\sin \left( \omega t\right) \\
-ih_{0}^{\text{sp}}\sin \left( \omega t\right) & \epsilon _{\text{s}}%
\end{array}%
\right) , \\
&J\left( t\right)=\left(
\begin{array}{cc}
t_{\text{p}}-ih_{1}^{\text{pp}}\sin \left( \omega t\right) & ih_{1}^{\text{sp
}}\sin \left( \omega t\right) \\
-ih_{1}^{\text{sp}}\sin \left( \omega t\right) & t_{\text{s}}-ih_{1}^{\text{
ss}}\sin \left( \omega t\right)%
\end{array}%
\right) ,
\end{align}
where $\epsilon _{\text{s}}$ and $\epsilon _{\text{p}}$ is the onsite
energy, $t_{\text{s}}$ and $t_{\text{p}}$ are the hopping amplitude from the
static part. $h_{0}^{sp}$ denotes shaking-induced on-site coupling between
the $s$- and $p$-band, and $h_{1}^{\text{pp}}$, $h_{1}^{\text{ss}}$, $h_{1}^{%
\text{sp}}$ denote shaking-induced nearest neighboring hopping within $s$%
-band and $p$-band, between $s$- and $p$-bands, respectively \cite{supple}.
For a given lattice depth $V$, $\epsilon _{\text{s}}$, $\epsilon _{\text{p}}$%
, $t_{\text{s}}$ and $t_{\text{p}}$ are fixed, and $h_{0}^{\text{sp}}$, $%
h_{1}^{\text{ss}}$, $h_{1}^{\text{pp}}$ and $h_{1}^{\text{sp}}$ scale
linearly with $k_{r}b$.

\begin{figure}[tbp]
\includegraphics[width=3.4in]
{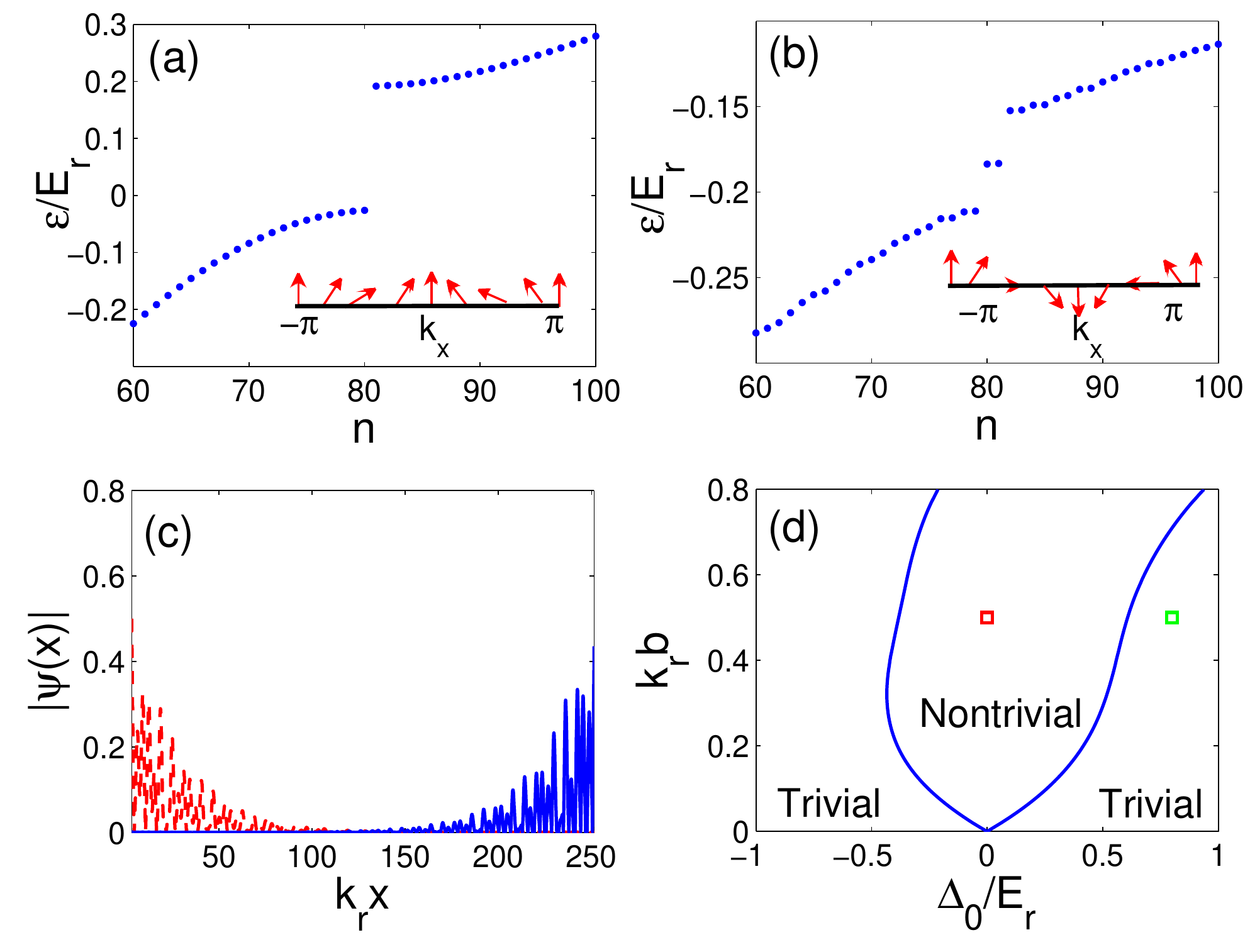}
\caption{ (a-b) Quasi-energy spectrum of a finite size one-dimensional
shaking optical lattice. $\Delta_0/E_\text{r}=0$ and $k_rb=0.5$ for (a); and
$\Delta_0/E_\text{r}= 0.8$ and $k_rb=0.5$ for (b), as marked in (d). Inset: winding of ${\bf B}(k_x)$ in $yz$ plane as $k_x$ changes from $-\pi$ to $\pi$. See text for definition of ${\bf B}(k_x)$.  (c):
The wave functions for the in-gap states of (b). (d): Phase diagram in terms
of shaking frequency $\Delta_0/E_\text{r}$ and shaking amplitude $k_rb$. $%
V=3E_\text{r}$ }
\label{fig2}
\end{figure}

With the Hamiltonian Eq. \ref{H1d} and \textbf{Method I}, we find phase
transitions between topological trivial and nontrivial phase, by changing
frequency via $\Delta _{0}=(\epsilon _{\text{p}}-\epsilon _{\text{s}%
}-2\omega)/2$ and shaking amplitude $k_r b$. A phase diagram is shown in
Fig. \ref{fig2}(d). The topological nontrivial state possess a pair of
in-gap states in the quasi-energy spectrum of a finite size lattice, as
shown in Fig.\ref{fig2}(b), whose corresponding wave functions are well
localized in the edges, see Fig. \ref{fig2}(c). In contrast, in the
topological trivial regime, there is no in-gap state in the quasi-energy
spectrum, see Fig.\ref{fig2}(a). As one can see clearly in Fig. \ref{fig2}%
(d), even for relatively small shaking amplitude $k_r b\approx 0.1$, there
is a quite large regime for topological nontrivial phase in the phase
diagram.

To understand the emergence of topological nontrivial phase, we write the
Hamiltonian into momentum space as $\hat{H}(t)=\sum_{k_{x}}\hat{\Psi}%
_{k_{x}}^{\dag }H_{k_{x}}\hat{\Psi}_{k_{x}}$ ($k_{x}$ in unit of $1/a$, $a$
is lattice spacing) and $H_{k_{x}}$ is given by
\begin{align}
& H_{k_{x}}=\left(
\begin{array}{cc}
\epsilon _{\mathrm{p}}+2t_{\mathrm{p}}\cos k_{x} & 0 \\
0 & \epsilon _{\mathrm{s}}+2t_{\mathrm{s}}\cos k_{x}%
\end{array}%
\right) +\sin \left( \omega t\right) \times   \notag \\
& \left(
\begin{array}{cc}
2h_{1}^{\mathrm{pp}}\sin k_{x} & i\left( h_{0}^{\mathrm{sp}}+2h_{1}^{\mathrm{%
\ sp}}\cos k_{x}\right)  \\
-i\left( h_{0}^{\mathrm{sp}}+2h_{1}^{\mathrm{sp}}\cos k_{x}\right)  &
2h_{1}^{\mathrm{ss}}\sin k_{x}%
\end{array}%
\right)
\end{align}%
With two-phonon resonance condition $2\omega \approx \epsilon _{\text{p}%
}-\epsilon _{\text{s}}$, $p$-band with dispersion $\epsilon _{\text{p}}+2t_{%
\text{p}}\cos (k_{x})$ and two-phonon dressed $s$-band with dispersion $%
\epsilon _{\text{s}}+2\omega +2t_{\text{s}}\cos (k_{x})$ form two close
bands as schematized in Fig. 1(a). Therefore, we shall first apply a unitary
rotation $O\left( t\right) =\exp \left( i\omega t\sigma _{z}\right) $ that
leads to
\begin{equation}
\hat{H}_{\mathrm{rot}}\left( t\right) =\hat{H}_{0}+\sum_{n=\pm 1,\pm 3}\hat{H%
}_{n}e^{in\omega t}.
\end{equation}%
Here $\hat{H}_{0}=\left( \Delta _{0}+2t\cos k_{x}\right) \sigma _{z}$, $\hat{%
H}_{1}=-ih_{1}\sin k_{x}\sigma _{z}-\left( h_{0}^{\mathrm{sp}}+2h_{1}^{%
\mathrm{sp}}\cos k_{x}\right) \sigma _{+}/2$, $\hat{H}_{3}=\left( h_{0}^{%
\mathrm{sp}}+2h_{1}^{\mathrm{sp}}\cos k_{x}\right) \sigma _{+}/2$, $\hat{H}%
_{-n}=\hat{H}_{n}^{\dag }$, where $2t=t_{\text{p}}-t_{\text{s}}$, $%
2h_{1}=h_{1}^{\text{pp}}-h_{1}^{\text{ss}}$. In the tight-binding regime, we
have $\omega \gg \Delta _{0},t,h_{0}^{\text{sp}},h_{1}^{\lambda \lambda }$.
Thus, it fulfills the condition to apply \textbf{Method II}. By choosing $%
T_{i}=T/4$ and following the formula Eq. \ref{Heff}, the effective
Hamiltonian can be deduced as $H_{\mathrm{eff}}=\mathbf{B}\left(
k_{x}\right) \cdot \mathbf{\sigma }$ where $B_{x}=0$ and
\begin{align}
& B_{y}=\frac{2\left( h_{0}^{\mathrm{sp}}+2h_{1}^{\mathrm{sp}}\cos
k_{x}\right) }{\omega }\left[ h_{1}\sin k_{x}-\frac{2\left( \Delta
_{0}+2t\cos k_{x}\right) }{3}\right]   \notag \\
& B_{z}=\Delta _{0}+2t\cos k_{x}+\frac{2\left( h_{0}^{\mathrm{sp}}+2h_{1}^{%
\mathrm{sp}}\cos k_{x}\right) ^{2}}{3\omega }.  \label{Bkx}
\end{align}%
This describes a momentum-dependent magnetic field in the $yz$ plane of the
Bloch sphere, which is analogous to momentum space representation of
Su-Schrieffer-Heeger model \cite{SSH}. Su-Schrieffer-Heeger model exhibits
topological nontrivial phase characterized by non-zero Zak phase \cite{Zak},
which has been realized and measured recently in double-well optical lattice
\cite{BlochSSH}. Whether the system is topologically trivial or not depends
on whether $\mathbf{B}(k_{x})$ has a nonzero winding number in the $yz$
plane as $k_{x}$ changes form $-\pi $ to $\pi $. As shown in the inset of
Fig. \ref{fig2}(a) and (b), for topological trivial case of Fig. \ref{fig2}%
(a), $\mathbf{B}(k_{x})$ has a winding number zero; while for topological
nontrivial case of Fig. \ref{fig2}(b), the winding number of $\mathbf{B}%
(k_{x})$ equals to one. From Eq. \ref{Bkx}, it is easy to see that when $%
|\Delta _{0}|$ is large enough, $B_{z}$ is dominated by the constant term
and therefore $\mathbf{B}(k_{x})$ has no winding. That explains why the
topological nontrivial phase occurs around $\Delta _{0}\approx 0$.

Before ending this part, it is worth to note that the two-phonon resonance
condition plays a crucial role here. In contrast, if we consider one-phonon
resonance condition $\omega \sim \epsilon _{\text{p}}-\epsilon _{\text{s}}$,
it is straightforward to show by similar analysis that there will be no
topological nontrivial phase \cite{supple}.

\textit{Two-Dimensional Case.} We employ the laser setup for a
two-dimensional honeycomb lattice used by ETH group \cite{Esslinger}, as
shown in Fig.\ref{fig1}(c). The interference of $X$ and $Y$ beams gives a
chequerboard of spacing $\lambda /\sqrt{2}$. $\bar{X}$ gives an additional
standing wave with spacing $\lambda /2$. When $V_{\bar{X}}\gg V_{Y}\gtrsim
V_{X}$, it leads to a honeycomb lattice as shown in Fig.\ref{fig1}(c). Using
the same method as in one-dimensional case, optical lattice can be shaken in
both $x$ and $y$ directions with a phase difference $\pi /2$. This gives
rise to a time-dependent potential are:
\begin{eqnarray*}
&&V(x,y,t)= \\
&&-V_{\bar{X}}\cos ^{2}\left[ k_{r}\left( x+b\cos \omega t\right) +\theta /2%
\right] \\
&&-V_{X}\cos ^{2}\left[ k_{r}\left( x+b\cos \omega t\right) \right]
-V_{Y}\cos ^{2}\left[ k_{r}\left( y+b\sin \omega t\right) \right] \\
&&-2\alpha \sqrt{V_{X}V_{Y}}\left[ \cos k_{r}\left( x+b\cos \omega t\right) %
\right] \cos \left[ k_{r}\left( y+b\sin \omega t\right) \right] .
\end{eqnarray*}%
Here $\theta $ controls the energy offset $M$ between AB sublattice. Similar
as in one-dimensional case, transferring into the comoving frame $%
x\rightarrow x+b\cos \left( \omega t\right) $, and $y\rightarrow y+b\sin
\left( \omega t\right) $, one obtains a Hamiltonian with time-dependent
vector potential term
\begin{equation}
H(t)=\frac{1}{2m}\left[ \mathbf{k}-\mathbf{A}(t)\right] ^{2}+V(x,y),
\end{equation}%
where $A_{x}(t)=m\omega b\sin (\omega t)$ and $A_{y}(t)=-m\omega b\cos
(\omega t)$. It is equivalent to an ac electrical field in the
two-dimensional plane $\mathbf{E}(t)=m\omega ^{2}b\left( \cos (\omega
t),\sin (\omega t)\right) $. With the tight-binding approximation and
Peierls substitution, the Hamiltonian is given by
\begin{equation}
\hat{H}(t)=\sum\limits_{\left\langle ij\right\rangle }\left( a_{A,j}^{\dag
},a_{B,j}^{\dag }\right) \left(
\begin{array}{cc}
M\delta _{ji} & t_{j}e^{i\mathbf{A}\left( t\right) \mathbf{\cdot d}_{ji}} \\
h.c. & -M\delta _{ji}%
\end{array}%
\right) \left(
\begin{array}{c}
a_{A,i} \\
a_{B,i}%
\end{array}%
\right)  \label{tight_binding_2d}
\end{equation}%
where $\mathbf{d}_{ji}$ is the vector from site $i$ pointing to site $j$, and $t_j$ is the hopping amplitude. 

\begin{figure}[tbp]
\includegraphics[ width=3.3in]
{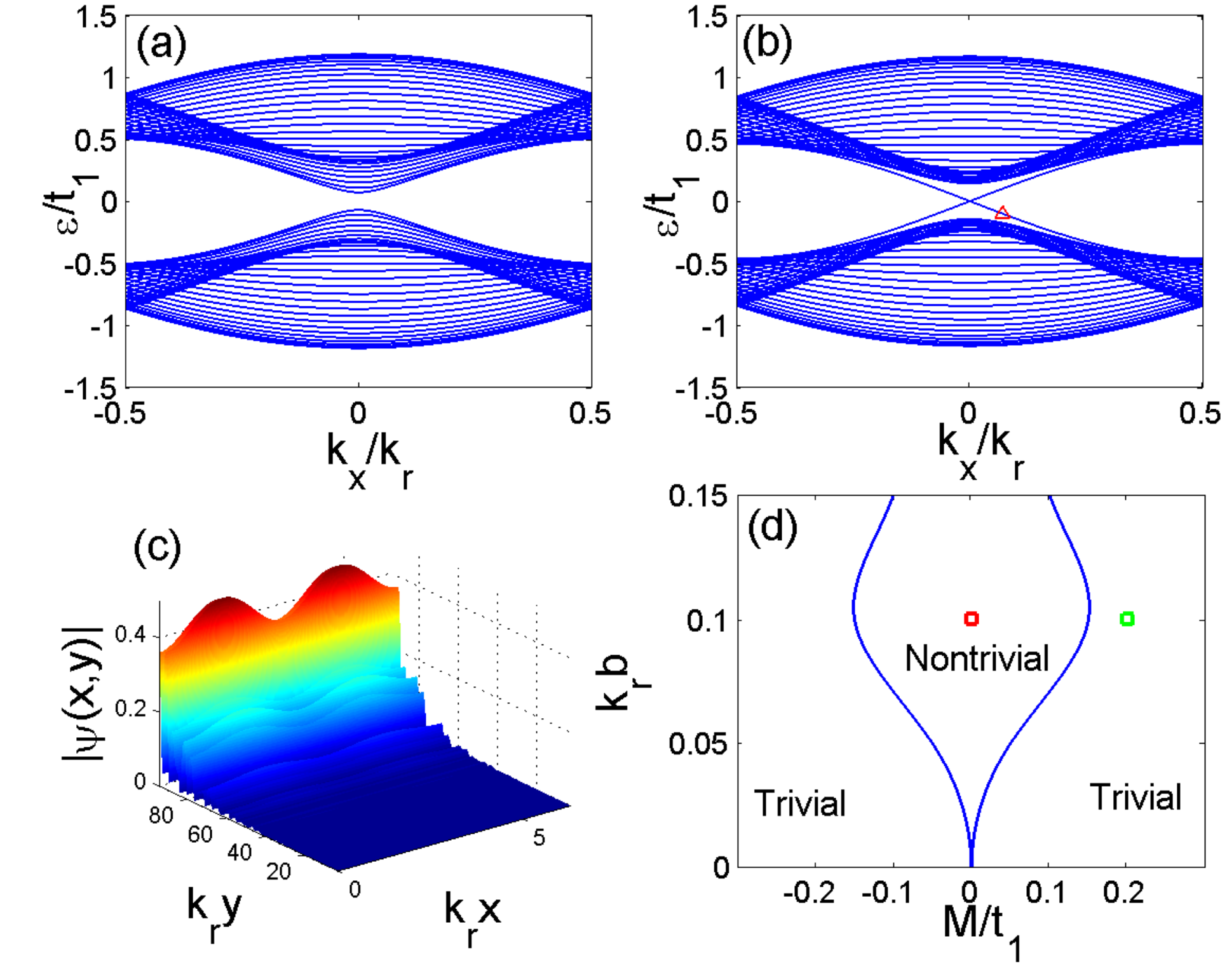}
\caption{ (a-b) Quasi-energy spectrum of a finite size two-dimensional
shaking honeycomb lattice with armchair edge. $M/t_{1}=0$ and $k_rb=0.1$ for
(a); and $M/t_{1}=0.3$ and $k_rb=0.1$ for (b), as marked in (d). (c): The
wave functions for the in-gap states of (b). (d): Phase diagram in terms of the on-site energy difference of AB sub lattice $M/E_\text{r}$ and shaking amplitude $k_rb$. $\hbar
\protect\omega/t_{1}$ is fixed at $6.28$. }
\label{fig3}
\end{figure}

Applying \textbf{Method I} to this model, we find a similar phase diagram
that contains topological trivial and nontrivial phases, as shown in Fig. %
\ref{fig3}(d). In this case, the phase diagram is controlled by parameter $%
M/E_\text{r}$ and shaking amplitude $k_rb$. Similarly, the topological
trivial phase has no in-gap states in quasi-energy spectrum (Fig. \ref{fig3}%
(a)), and topological nontrivial phase has a pair in-gap states (Fig. \ref%
{fig3}(b)), whose corresponding wave function (Fig. \ref{fig3}(c)) is
localized at the edge of the two-dimensional sample. Same as one-dimensional
case, even for small shaking amplitude of $k_r b\approx 0.1$, the
topological nontrivial regime occupies a large parameter space.

To illustrate the relation of this topological nontrivial phase with the
Haldane model and the quantum anomalous Hall effect, we first expand
Hamiltonian Eq. \ref{tight_binding_2d} as $H\left( \mathbf{k},t\right)
=\sum_{n=-\infty }^{\infty }\hat{H}_{n}\left( \mathbf{k}\right) e^{in\omega
t}$. $\hat{H}_{0}$ gives rise to a static honeycomb lattice structure, which
contains two bands with band-width $\sim 2t_{j}$ and band-gap $\sim 2M$.
When $\omega \gg 2M,2t_{j}$, the condition for applying \textbf{Method II}
is satisfied, and it yields an effective Hamiltonian $H_\text{eff}(\mathbf{k})
=\mathbf{B}(\mathbf{k})\cdot\mathbf{\sigma}$. The explicit form of $\mathbf{B}(%
\mathbf{k})$ is given in supplementary material \cite{supple}. This
effective Hamiltonian can be compared with Haldane model. If $\mathbf{B}(%
\mathbf{k})$ fully covers the Bloch sphere as $\mathbf{k}$ goes over the
Brillouin zone, this phase is topologically nontrivial and exhibits quantum
anomalous Hall effect \cite{Haldane}.

For small shaking amplitude, at the leading order of $k_rb$, $B_x(\mathbf{k}%
) $ and $B_y(\mathbf{k})$ are given by the static part of the honeycomb
lattice Hamiltonian. Due to the Dirac point structure, $\{B_x, B_y\}$ has
desired winding structure in the $xy$ plane. $B_z(\mathbf{k})$ can be
written as $M+D(\mathbf{k})$, and for small shaking amplitude, $D(\mathbf{k}%
) $ scales linearly with $k_r b$. If $|M|>D(\mathbf{k})$ for all $\mathbf{k}$%
, either due to small $k_r b$ or large $|M|$, $B_z$ always has the same sign
as $M$ and therefore spin can only point to half of the Bloch sphere, the
resulting state will still be topological trivial, as shown in Fig. \ref%
{fig3}(d).

As $k_rb$ increases, $D(\mathbf{k})$ will become larger than $M$ in certain
regime of $\mathbf{k}$ space. In particular, for our model, similar as the
case of Haldane mode, $D(\mathbf{k})$ takes opposite sign between two Dirac
points (where both $B_x$ and $B_y$ vanish), and its absolute value is larger
than $|M|$. Thus, $B_z$ takes opposite values between two Dirac points and
the spin vector points to north and south poles, respectively, at two Dirac
points. This feature, together with nontrivial winding of $\{B_x,B_y\}$ in
the $xy$ plane, gives rise to a topologically nontrivial coverage of spin
vector in the Bloch sphere. Consequently, it enters topological nontrivial
phase, with a non-zero Chern number and chiral edge state, as shown in Fig. %
\ref{fig3}. With noninteracting fermions in this setup, it will exhibit
quantum anomalous Hall effect with quantized Hall conductance, which can be
measured by various methods \cite{AQH_cold_atom,Hall}.

\textit{Final Remark:} We believe the schemes and examples presented in this
work open a new route toward topological states in cold atom systems. It
will be more interesting to generalize the current work to three-dimension
and the case with interactions.

\textit{Acknowledgment}: This work is supported by Tsinghua University
Initiative Scientific Research Program, NSFC Grant No. 11174176, and
NKBRSFC under Grant No. 2011CB921500.

\begin{widetext}

\section{Supplementary Materials}

\subsection{Definition of Parameters}

The shaking induced hopping parameters in Eq. (6) of main text are defined as%
\begin{eqnarray*}
h_{0}^{\mathrm{sp}} &=&b\omega \int dx\phi _{\mathrm{s}}\left( x\right)
\frac{\partial }{\partial x}\phi _{\mathrm{p}}\left( x\right) , \\
h_{1}^{\mathrm{sp}} &=&b\omega \int dx\phi _{\mathrm{s}}\left( x-a\right)
\frac{\partial }{\partial x}\phi _{\mathrm{p}}\left( x\right) , \\
h_{1}^{\mathrm{ss}} &=&b\omega \int dx\phi _{\mathrm{s}}\left( x-a\right)
\frac{\partial }{\partial x}\phi _{\mathrm{s}}\left( x\right) , \\
h_{1}^{\mathrm{pp}} &=&b\omega \int dx\phi _{\mathrm{p}}\left( x-a\right)
\frac{\partial }{\partial x}\phi _{\mathrm{p}}\left( x\right) .
\end{eqnarray*}%
Here $\phi _{\mathrm{s}}\left( x\right) $ and $\phi _{\mathrm{p}}\left(
x\right) $ are the wave functions of s-orbit and p-orbit. $a=\pi /k_{r}$ is
the lattice constant in one dimensional optical lattice.

\subsection{One-Phonon Transition}

Similarly, we shall first apply a unitary rotation $O(t)=\exp \left( i\omega
t\sigma _{z}/2\right) $ to make $p$-band and one-phonon dressed $s$-band
nearly degenerate. Then, following formula we reach an effective Hamiltonian
as:%
\begin{eqnarray}
H_{\mathrm{eff}} &=&\left[ \Delta _{0}+\frac{\omega }{2}+2t\cos k_{x}-\frac{%
\left( h_{0}^{\mathrm{sp}}+2h_{1}^{\mathrm{sp}}\cos k_{x}\right) ^{2}}{%
8\omega }\right] \sigma _{z}  \notag \\
&&+\frac{h_{0}^{\mathrm{sp}}+2h_{1}^{\mathrm{sp}}\cos k_{x}}{\omega }\left(
\Delta _{0}+\frac{3\omega }{2}+2t\cos k_{x}\right) \sigma _{y}
\end{eqnarray}%
It is obviously that the constant term in $B_{y}$ is about $\Delta
_{0}+3\omega /2\sim \omega $, which dominates the $k_{x}$-dependent terms, $%
\omega \gg t$. In this situation, $B_{y}$ cannot change the sign as $k_{x}$
changes from $-\pi $ to $\pi $, and gives zero winding. So it is in the
topological trivial phase. As mentioned previously, this is confirmed by the
\textbf{Method I}, where no edge state is found in the one-phonon resonance
regime.

\subsection{The effective Hamiltonian of shaked honeycomb lattice}

Transforming the Hamiltonian (12) in the main text into the momentum space,
one obtains:%
\begin{equation}
\hat{H}\left( t\right) =\sum\limits_{\mathbf{k}}\left( a_{A}^{\dag }\left(
\mathbf{k}\right) ,a_{B}^{\dag }\left( \mathbf{k}\right) \right) H\left(
\mathbf{k},t\right) \left(
\begin{array}{c}
a_{A}\left( \mathbf{k}\right) \\
a_{B}\left( \mathbf{k}\right)%
\end{array}%
\right) ,
\end{equation}%
where
\begin{equation*}
H\left( \mathbf{k},t\right) =\left(
\begin{array}{cc}
M & t_{1}e^{-i\left[ k_{x}-A_{x}\left( t\right) \right] a_{1}}+t_{2}e^{i%
\left[ k_{x}-A_{x}\left( t\right) \right] a_{2}+i\left[ k_{y}-A_{y}\left(
t\right) \right] a}+t_{2}e^{i\left[ k_{x}-A_{x}\left( t\right) \right]
a_{2}-i\left[ k_{x}-A_{x}\left( t\right) \right] a} \\
h.c. & -M%
\end{array}%
\right) ,
\end{equation*}%
Employing the Bessel functions, one can expanded this Hamiltonian as $%
H\left( \mathbf{k,}t\right) =\sum\limits_{n=-\infty }^{\infty }H_{n}\left(
\mathbf{k}\right) e^{in\omega t}$. For simplicity, we only
keep to $n=0,\pm 1$ terms. Such an approximation is valid for the small
shaking amplitude. Then one obtains
\begin{eqnarray*}
H_{0}\left( \mathbf{k}\right) &=&M\sigma _{z}+B_{x}\left( \mathbf{k}\right)
\sigma _{x}+B_{y}\left( \mathbf{k}\right) \sigma _{y} \\
H_{1}\left( \mathbf{k}\right) &=&-\left[ u_{x}\left( \mathbf{k}\right)
+iv_{x}\left( \mathbf{k}\right) \right] \sigma _{x}+\left[ u_{y}\left(
\mathbf{k}\right) +iv_{y}\left( \mathbf{k}\right) \right] \sigma _{y}
\end{eqnarray*}%
and $H_{-1}\left( \mathbf{k}\right) =H_{1}^{\dag }\left( \mathbf{k}\right) $%
, where
\begin{eqnarray*}
B_{x}\left( \mathbf{k}\right) &=&t_{1}^{\prime }\sin \left(
k_{x}a_{1}\right) -t_{2}^{\prime }\sin \left( k_{y}a+k_{x}a_{2}\right)
+t_{2}^{\prime }\sin \left( k_{y}a-k_{x}a_{2}\right) , \\
B_{y}\left( \mathbf{k}\right) &=&t_{1}^{\prime }\cos \left(
k_{x}a_{1}\right) +t_{2}^{\prime }\cos \left( k_{y}a+k_{x}a_{2}\right)
+t_{2}^{\prime }\cos \left( k_{y}a-k_{x}a_{2}\right) , \\
u_{x}\left( \mathbf{k}\right) &=&2t_{21}^{\prime \prime }\sin \left(
k_{y}a\right) \cos \left( k_{x}a_{2}\right) , \\
u_{y}\left( \mathbf{k}\right) &=&2t_{21}^{\prime \prime }\sin \left(
k_{y}a\right) \sin \left( k_{x}a_{2}\right) , \\
v_{x}\left( \mathbf{k}\right) &=&t_{1}^{\prime \prime }\sin \left(
k_{x}a_{1}\right) +2t_{22}^{\prime \prime }\cos \left( k_{y}a\right) \sin
\left( k_{x}a_{2}\right) , \\
v_{y}\left( \mathbf{k}\right) &=&t_{1}^{\prime \prime }\cos \left(
k_{x}a_{1}\right) -2t_{22}^{\prime \prime }\cos \left( k_{y}a\right) \cos
\left( k_{x}a_{2}\right) .
\end{eqnarray*}%
and%
\begin{eqnarray*}
t_{1}^{\prime } &=&t_{1}J_{0}\left( a_{1}m\omega b\right) , \\
t_{2}^{\prime } &=&t_{2}J_{0}\left( \sqrt{a^{2}+a_{2}^{2}}m\omega b\right) ,
\\
t_{1}^{\prime \prime } &=&t_{1}J_{1}\left( a_{1}\omega b/2\right) , \\
t_{21}^{\prime \prime } &=&\frac{a}{\sqrt{a^{2}+a_{2}^{2}}}t_{2}J_{1}\left(
\sqrt{a^{2}+a_{2}^{2}}m\omega b\right) , \\
t_{22}^{\prime \prime } &=&\frac{a_{2}}{\sqrt{a^{2}+a_{2}^{2}}}%
t_{2}J_{1}\left( \sqrt{a^{2}+a_{2}^{2}}m\omega b\right) .
\end{eqnarray*}%
Using Eq. (3) one can obtained the effective Hamiltonian (13)
\begin{eqnarray*}
H_{\mathrm{eff}}\left( \mathbf{k}\right) &=&H_{0}+\frac{\left[ H_{1},H_{-1}%
\right] }{\omega }-\frac{\left[ H_{1},H_{0}\right] }{\omega }+\frac{\left[
H_{-1},H_{0}\right] }{\omega } \\
&=&\left[ M+D\left( \mathbf{k}\right) \right] \sigma _{z}+B_{x}\left(
\mathbf{k}\right) \sigma _{x}+B_{y}\left( \mathbf{k}\right) \sigma _{y},
\end{eqnarray*}%
where
\begin{equation*}
D\left( \mathbf{k}\right) =-\frac{4}{\omega }\left[ u_{x}\left( \mathbf{k}%
\right) v_{y}\left( \mathbf{k}\right) -u_{y}\left( \mathbf{k}\right)
v_{x}\left( \mathbf{k}\right) +v_{x}\left( \mathbf{k}\right) B_{y}\left(
\mathbf{k}\right) +v_{y}\left( \mathbf{k}\right) B_{x}\left( \mathbf{k}%
\right) \right]
\end{equation*}%
Here for convenience we choose $\alpha =0$. One notes that $D\left( \mathbf{k%
}\right) $ breaks the time reversal symmetry, and $D^{\ast }\left( -\mathbf{k}%
\right) =-D\left( \mathbf{k}\right) $.

\end{widetext}


\begin{thebibliography}{99}
\bibitem{FTI} %Proposal of Floquent TI
N. H. Lindner, G. Refael, and V. Galitski, Nature Phys. \textbf{7}, 490
(2011).

\bibitem{Demler} T. Kitagawa, E. Berg, M. Rudner, and E. Demler, Phys. Rev.
B \textbf{82}, 235114 (2010)

\bibitem{Phonoic} %Realization of FTI in photonics crystal
M. C. Rechtsman, J. M. Zeuner, Y. Plotnik, Y. Lumer, D. Podolsky, F.
Dreisow, S. Nolte, M. Segev, and A. Szameit, Nature \textbf{496}, 196 (2013).

\bibitem{Bloch} M. Aidelsburger, M. Atala, S. Nascimbue, S. Trotzky, Y.-A.
Chen, and I. Bloch, Phys. Rev. Lett. \textbf{107}, 255301 (2011)

\bibitem{spielman} K. Jimez-Garc, L. J. LeBlanc, R. A. Williams, M. C.
Beeler, A. R. Perry, and I. B. Spielman, Phys. Rev. Lett. \textbf{108},
225303 (2012)

\bibitem{Bloch2013} M. Aidelsburger, M. Atala, M. Lohse, J. T. Barreiro, B.
Paredes, I. Bloch, arXiv: 1308.0321

\bibitem{Ketterle2013} H. Miyake, G. A. Siviloglou, C. J. Kennedy, W. Cody
Burton, W. Ketterle, arXiv: 1308.1431

\bibitem{Sengstock1} %Sengstock's experiment on shaking triangular lattice
J. Struck, C. \"Olschl\"ager, R. Le Targat, P. Soltan-Panahi, A. Eckardt, M.
Lewenstein, P. Windpassinger, and K. Sengstock, Science \textbf{333}, 996
(2011).

\bibitem{Sengstock2}
%Sengstock's experiment on shaking to generate gauge field
J. Struck, C. \"Olschl\"ager, M. Weinberg, P. Hauke, J. Simonet, A. Eckardt,
M. Lewenstein, K. Sengstock, and P. Windpassinger, Phys. Rev. Lett. \textbf{%
108}, 225304 (2012).

\bibitem{Cheng} C. V. Parker, L. C. Ha, and C. Chin, arXiv: 1305.5487 (2013).

\bibitem{SSH} %SSH model
W. P. Su, J. R. Schrieffer, and A. J. Heeger, Phys. Rev. Lett. \textbf{42},
1698 (1979).

\bibitem{Esslinger} %ETH dirac point
L. Tarruell, D. Greif, T. Uehlinger, G. Jotzu, and T. Esslinger, Nature
\textbf{483}, 302 (2012).

\bibitem{Haldane} F. D. M. Haldane, Phys. Rev. Lett. \textbf{61}, 2015 (1988)

\bibitem{AQH} %Xue's AQH
C.-Z. Chang, J. Zhang, X. Feng, J. Shen, Z. Zhang, M. Guo, K. Li, Y. Ou, P.
Wei, L.-L. Wang, Z.-Q. Ji, Y. Feng, S. Ji, X. Chen, J. Jia, X. Dai, Z. Fang,
S.-C. Zhang, K. He, Y. Wang, L. Lu, X.-C. Ma, and Q.-K. Xue, Science \textbf{%
340}, 167 (2013).

\bibitem{AQH_cold_atom} L. B. Shao, Shi-Liang Zhu, L. Sheng, D. Y. Xing, and
Z. D. Wang, Phys. Rev. Lett. \textbf{101}, 246810 (2008);

\bibitem{AQH_cold_atom2} X. J. Liu, X. Liu, C. Wu, and J. Sinova, Phys. Rev.
A \textbf{81}, 033622 (2010); M. Zhang, H. H. Hung, C. Zhang, and C. Wu,
Phys. Rev. A \textbf{83}, 023615 (2011);

\bibitem{supple} See supplementary materials for (i) explicit definition of
parameters in the one-dimensional model; (ii) one-phonon transition regime
of one-dimensional optical lattice case and (iii) the explicit definition
for $\mathbf{B}(\mathbf{k})$ for two-dimensional honeycomb model.

\bibitem{Zak} J. Zak, Phys. Rev. Lett. \textbf{62}, 2747 (1989)

\bibitem{BlochSSH} M. Atala, M. Aidelsuburger, J. T. Barreiro, D. Abanin, T.
Kitagawa, E. Demler, and I. Blcoh, Nature Phy. \textbf{9}, 795 (2013)

\bibitem{Hall} R. O. Umucallar, H. Zhai, and M. \"O. Oktel, Phys. Rev. Lett.
\textbf{100}, 070402 (2008); E. Zhao, N. Bray-Ali, C. J. Williams, I. B.
Spielman, and I. I. Satija, Phys. Rev. A \textbf{84}, 063629 (2011); N.
Goldman, J. Beugnon, and F. Gerbier, Phys. Rev. Lett. \textbf{108}, 255303
(2012)
\end{thebibliography}
\end{document}